\documentclass[a4paper,fleqn,usenatbib,useAMS]{mnras}

\usepackage{graphicx}	
\usepackage{amsmath}	
\usepackage{amssymb}	

\usepackage[T1]{fontenc}
\usepackage{aecompl}

\usepackage{newtxtext,newtxmath}

\newcommand*\fermi{\textit{Fermi}}
\newcommand*\hst{\textit{HST}}

\title[Timing analysis of the Vela pulsar from Iqueye observations]{Timing analysis and pulse profile of the Vela pulsar in the optical band from Iqueye observations}

\author[Spolon et al.]{A. Spolon$^{1}$, L. Zampieri$^{2}$\thanks{E-mail: \href{mailto:luca.zampieri@oapd.inaf.it}{luca.zampieri@inaf.it}}, A. Burtovoi$^{2}$, G. Naletto$^{1,2}$, C. Barbieri$^{1,2}$, M. Barbieri$^{3}$, 
\newauthor A. Patruno$^{4,5}$ and E. Verroi$^{6}$
\\
$^{1}$Department of Physics and Astronomy ``G. Galilei'', University of Padova, Via Marzolo, 8, I-35131 Padova, Italy\\
$^{2}$INAF - Astronomical Observatory of Padova, Vicolo dell' Osservatorio 5, I-35122 Padova, Italy\\
$^{3}$Department of Physics, University of Atacama, Copayapu 485, Copiapo, Chile\\
$^{4}$Leiden Observatory, Leiden University, Neils Bohrweg 2, 2333 CA Leiden, The Netherlands\\
$^{5}$ASTRON, the Netherlands Institute for Radio Astronomy, Postbus 2, 7900 AA Dwingeloo, The Netherlands\\
$^{6}$TIFPA - Trento Institute for Fundamentals Physics Applications, Via Sommarive 14, I-38123 Povo, Italy\\
}

\pubyear{2017}

\begin{document}
\label{firstpage}
\pagerange{\pageref{firstpage}--\pageref{lastpage}}
\maketitle

\begin{abstract}
The Vela pulsar is among a number of pulsars which show detectable optical pulsations. We performed optical observations of this pulsar in January and December 2009 with the Iqueye instrument mounted at the ESO 3.5 m New Technology Telescope. Our aim was to perform phase fitting of the Iqueye data, and to measure the optical pulse profile of the Vela pulsar at high time resolution, its absolute phase and rotational period. We calculated for the first time an independent optical timing solution and obtained the most detailed optical pulse profile available to date. Iqueye detected a distinct narrow component on the top of one of the two main optical peaks, which was not resolved in previous observations, and a third statistically significant optical peak not aligned with the radio one. The quality of the Iqueye data allowed us to determine the relative time of arrival of the radio-optical-gamma-ray peaks with an accuracy of a fraction of a millisecond. We compare the shape of the Iqueye pulse profile with that observed in other energy bands and discuss its complex multi-wavelength structure.
\end{abstract}


\begingroup
\let\clearpage\relax
\endgroup
\newpage

\section{Introduction}\label{sec:1}
The Vela pulsar (PSR J0835$-$4510) is one of the most intense sources in the gamma-ray and radio bands. It has a characteristic age of about 11 kyr \citep{Lyne1996} and is located in the southern sky at a distance of 290 pc \citep{Caraveo2001}. Unlike the majority of the known pulsars, it also shows pulsed optical emission at detectable levels, with an optical magnitude of V$\sim$23.6 \citep{Mignani2001}.

PSR J0835$-$4510 was discovered during a survey looking for pulsars in the southern sky \citep{Large1968}. It shows a periodicity of 89.3 ms and, at the time of discovery, it was the pulsar with the shortest rotational period. \citet{Large1968} noted that the pulsar was located in the center of the Vela nebula, which has a diameter between 4$^{\circ}$ and 5$^{\circ}$, already known to be a radio source. Compared to the Crab pulsar (see e.g. \citealt{Oosterbroek2008, Germana2012, Zampieri2014}), the source is much weaker in the optical band and optical pulses were discovered only after a prolonged observing campaign using a dual photometer mounted on the Anglo-Australian telescope \citep{Wallace1977}.
Later on, optical pulsations of the Vela pulsar were detected by \citet{Manchester1978, Peterson1978, Manchester1980, Gouiffes1998}. Ultraviolet (UV) observations have been carried out with the Hubble Space Telescope (\hst) by \citealt{Romani2005}), who obtained the light curves and phase-resolved spectra in the near and far UV bands. The Vela pulsar presents also strong X-ray and gamma-ray pulsed emission with pulse profiles very different from those detected in the radio and optical bands (see e.g. \citealt{Manchester1977, Harding2002, Abdo2009_4, Kuiper2015}). The detection of X-ray pulsations from the Vela pulsar was reported for the first time by \citet{Harnden1972} and then by \cite{Harnden1973}. Detailed studies of the X-ray pulse profile have been done using the data from the ROSAT \citep{Oegelman1993, Seward2000}, \textit{Chandra} \citep{Helfand2001, Sanwal2002}, {\it Rossi-XTE} \citep{Strickman1999, Harding2002} and \textit{XMM-Newton} \citep{Manzali2007} telescopes. Thanks to the high time resolution of {\it Rossi-XTE}, \citet{Harding2002} obtained the most detailed light curve of the Vela pulsar in the X-ray band (2-30 keV) to date, finding that one of the main X-ray peaks has a double structure. Later, \citet{Kuiper2015} reported updated X-ray and soft gamma-ray pulse profiles. They re-analyzed larger data sets from different telescopes (\textit{XMM-Newton}, \textit{Rossi-XTE}, \textit{INTEGRAL} and \textit{COMPTEL}). Other instruments observed the Vela pulsar at gamma-ray energies (\textit{SAS-2}, \citealt{Thompson1975, Thompson1977}; \textit{COS-B}, \citealt{Bennett1977, Buccheri1978, Kanbach1980}; EGRET, \citealt{Kanbach1994}). The latest gamma-ray studies performed with the AGILE and \fermi-LAT instruments \citep{Pellizzoni2009, Abdo2009_4, Abdo2010_3} showed the complex energy-dependent structure of the Vela pulsar light curve at GeV energies, which consists of two main peaks with a third peak and/or bridge emission in between them. Recently, \citet{Djannati2017_gamma2016} reported the first detection of the pulsed emission from the Vela pulsar at 20-120 GeV with the H.E.S.S. Cherenkov telescope.

Studying optical pulsars within the framework of their multi-wavelength emission properties is crucial to improve our understanding of their emission mechanism. Multi-wavelength investigations of the pulsed emission are very important for testing and constraining theoretical models describing the structure of the magnetic field and location of the pulsar emitting regions. 

In this paper, we present results of the timing analysis of the Vela pulsar from the high time resolution optical observations carried out with the Iqueye instrument. Details on the observations and the data analysis procedure are reported in Sect. \ref{sec:2}. Results on the timing analysis and phase folding are presented in Sect. \ref{sec:3}. The discussion and conclusions follow in Sect. \ref{sec:4}.

\section{Observations and Data Reduction}\label{sec:2}
In this work we analyze the observations of the Vela pulsar taken with the single photon-counting photometer Iqueye mounted at the ESO 3.5 m New Technology Telescope (NTT) in two observational runs during 2009 January (from 19 through 20) and December (from 14 through 18). Iqueye is an extremely high speed optical photon counter based on single photon avalanche photodiodes (SPAD) silicon detectors, which are equipped with a digital acquisition system that records arrival times of each single photon (for details see \citealt{Naletto2009}).

A log of the Iqueye observations of the Vela pulsar performed in two 2009 runs at NTT is reported in Table \ref{tab:osservazioni}. Apart from a couple of very short exposures of less than 100 s, taken to test the instrument response, all the other observations have duration longer than 5 minutes, with most of them being above 30 minutes.
 
The data reduction is performed converting the time tags of each photon in UTC by means of a dedicated software\footnote{\texttt{QUEST} v. 1.1.5, Zoccarato, P. 2015, Internal Technical Report; see also \citealt{Zampieri2015}}.
Then, the photon arrival times are barycentered by means of the software \texttt{TEMPO2},
which adopts the Barycentric Coordinate Time and is accurate at the 1 ns level of precision \citep{Hobbs2006,Edwards2006}.

\begin{figure}
	\centering
	\includegraphics[width=0.5\textwidth]{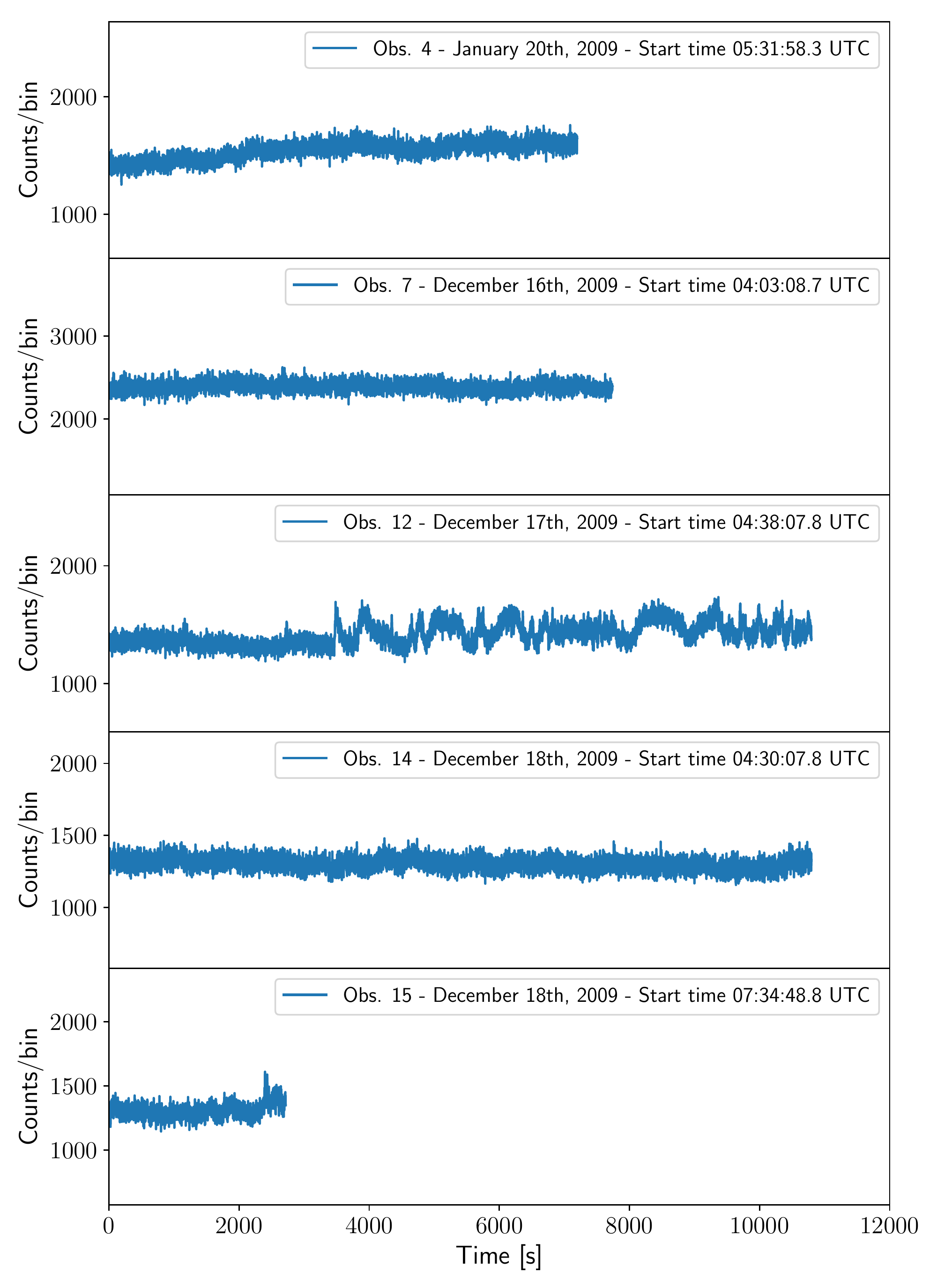}
	\caption{Light curves with time bin of 1 s of the 5 observations with the most significant signal (see Table \ref{tab:osservazioni} and the text for details).}
	\label{fig:lc1uk}
\end{figure}

In order to check the quality of the observations, we initially calculate light curves with bins of 1 s and inspect them. In Fig. \ref {fig:lc1uk} we show the light curves of the 5 observations with the lowest background.
Obs. 4 and 14 (see Table \ref{tab:osservazioni}) have the cleanest light curves.
The total count rate measured by all four Iqueye detectors is about 1300 counts s$^{-1}$. Obs. 7 shows a comparatively higher background level.
Obs. 12 has a good signal at the beginning but, as time goes on, the background becomes higher because of the passage of significant veils that increase the amount of scattered light.
Obs. 15 has the shortest duration. The rise of the count rate towards the end of the acquisition is caused by the passage of thin clouds.
As shown in Appendix \ref{app:1}, in all these low-background-contaminated observations (obs. 4, 7, 12, 14 and 15) we detected the pulsar signal.

\begin{table*}
\centering
\caption{Log of the 2009 observations of the Vela pulsar taken with the Iqueye instrument at the ESO 3.5 m  NTT telescope in Chile.}
\label{tab:osservazioni}
\begin{tabular}{clrlcr}
\hline
\hline
\noalign{\smallskip}
 Obs.     & \multicolumn{1}{c}{Observation ID} & \multicolumn{2}{c}{Start time (UTC)} & \multicolumn{1}{c}{Start time [MJD]} & \multicolumn{1}{c}{Duration [s]}	\\
\noalign{\smallskip}
\hline
\noalign{\smallskip}
1 & 20090119-070620UTC & January 19 & 07:16:15.8 & 54850.302960 & 72.0	\\ 
2 & 20090119-071240UTC & January 19 & 07:18:03.8 & 54850.304210 & 878.1	\\
3 & 20090119-073755UTC &  January 19  & 07:43:23.8 & 54850.321803 & 3617.0	\\ 
4 & 20090120-052536UTC &  January 20 & 05:31:58.3 & 54851.230534 & 7207.1	\\ 
5 & 20091214-081325UTC &  December 14  & 08:16:21.4 & 55179.344692 & 1264.0	\\ 
6 & 20091216-035420UTC &  December 16 & 04:01:03.7 & 55181.167403 & 5.0	\\ 
7 & 20091216-040003UTC & December 16 & 04:03:08.7 & 55181.168850 & 7740.2	\\ 
8 & 20091216-060939UTC & December 16& 06:14:09.1 & 55181.259827 & 301.0	 \\ 
9 & 20091216-064816UTC & December 16 & 06:51:19.2 & 55181.285638 & 3601.0	\\ 
10 & 20091217-034741UTC & December 17 & 03:51:06.7 & 55182.160494 & 301.0	\\ 
11 & 20091217-040317UTC & December  17& 04:06:44.0 & 55182.171342 & 1827.0	 \\ 
12 & 20091217-043458UTC & December 17 & 04:38:07.8 & 55182.193145 & 10798.8	\\ 
13 & 20091217-073602UTC & December 17& 07:39:53.3 & 55182.319368 & 2693.6	\\ 
14 & 20091218-042650UTC & December 18& 04:30:07.8 & 55183.187590 & 10798.3	\\ 
15 & 20091218-073137UTC & December 18  & 07:34:48.4 & 55183.315835 & 2716.8	\\ 
\noalign{\smallskip}
\hline
\noalign{\smallskip}
\end{tabular}
\end{table*} 

\section{Analysis and Results}\label{sec:3}
\subsection{Phase fitting of the Iqueye data}\label{sec:3.1}
We started our analysis performing a phase fitting of the Iqueye data acquired during the 2009 December run. Following the approach already presented in \citet{Germana2012} and \citet{Zampieri2014}, we divided observations 7, 12, 14, 15 in several segments and folded them over a reference period $P_{\rm init}$. Each segment is sufficiently long ($\sim$ 45 min) to detect the main peaks of the pulse profile. The reference period $P_{\rm init}$ is equal to that measured from the Iqueye data of Obs. 12 ($P_{\rm init}$$=$0.08936694 s; see Appendix \ref{app:1}). Each segment is binned with 50 bins per period\footnote{For Obs. 7 we used 41 phase bins because of the lower signal-to-noise ratio.}.

\begin{figure}
	\centering
	\includegraphics[width=0.5\textwidth]{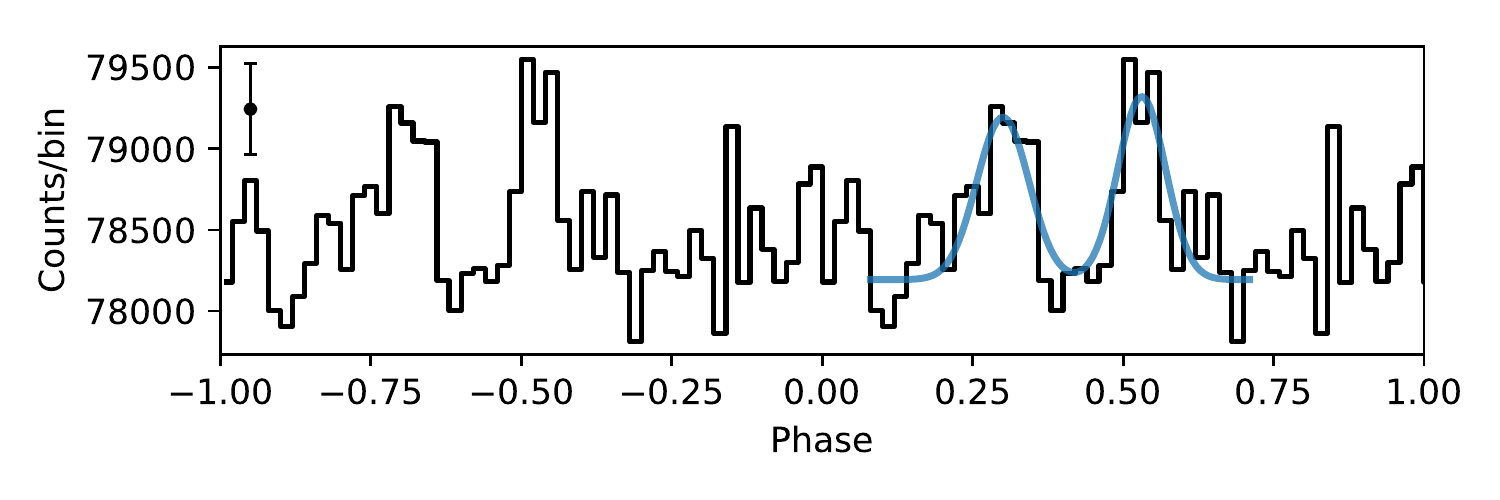} \\
	\caption{One of the 45 min segments of Obs. 12 folded with the period $P_{\rm init}$$=$0.08936694 s. The blue line shows two Gaussian functions used to fit the main optical peaks. The average 1$\sigma$ error bar is shown on the top left.}
	\label{fig:Obs12_segm}
\end{figure}

In order to determine the phase of the Vela pulsar, we fitted the two main peaks in each segment of the Iqueye pulse profile with the sum of two Gaussian functions plus a constant (see Fig. \ref{fig:Obs12_segm}). The measured rotational phase ($\psi_{\rm obs}(t)$) is the mean value of the Gaussian corresponding to the second and most prominent peak. To improve the accuracy of the measurement, we fixed the separation ($\Delta\phi_{\rm 12}$) and widths ($\sigma_1$ and $\sigma_2$) of the two peaks at the following values: $\Delta\phi_{\rm 12} = 0.231$, $\sigma_1 = 0.040$ and $\sigma_2 = 0.043$. They were obtained by fitting the two main peaks of the total Iqueye pulse profile (folded using the \fermi-LAT timing solution; see Section \ref{sec:3.2}) with the the sum of two Gaussian functions plus a constant. The typical 1$\sigma$ uncertainty on the position of the peak in each segment is 0.0071 in phase (or 0.63 ms).

\begin{figure}
	\centering
	\includegraphics[width=0.5\textwidth]{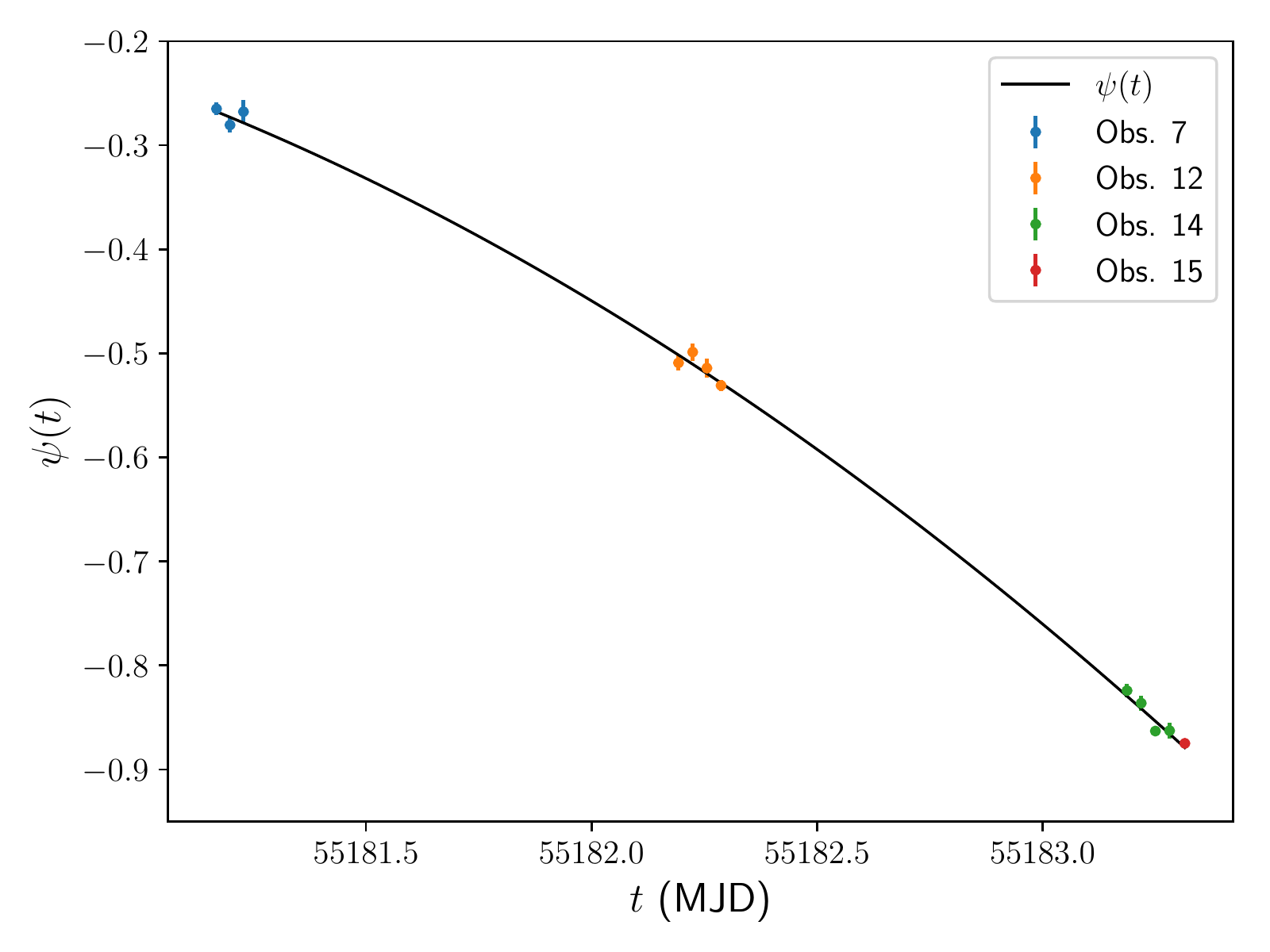}
	\caption{Phase drift of the Vela pulsar $\psi(t)$ with respect to uniform rotation. Time $t$ expressed in MJD.}
	\label{fig:Iqueye_drift}
\end{figure}

The measured peak position drifts in phase because of the pulsar spin-down. The drift with respect to uniform rotation with period $P_{\rm init}$ can be written as (\citealt{Germana2012,Zampieri2014})\footnote{We note, in fact, that following \citet{Germana2012} and \citet{Zampieri2014} $\psi(t)$ in equation~(\ref{eq:psi}) is defined as the opposite value of the drift.}:
\begin{equation}
	\psi(t) = \phi(t) - (t-t_0)/P_{\rm init} \, ,
	\label{eq:psi}
\end{equation}
where $t_0$ is a reference time set equal to MJD 55182 and $\phi(t)$ is the phase of the pulsar peak at time $t$. For a three-day baseline, the pulsar spin-down $\phi(t)$ can be reasonably well modelled with a second-order polynomial function:
\begin{equation}
	\phi(t) = \phi_0 + \nu_0(t-t_0) + \frac{\dot\nu_0}{2}(t-t_0)^2,
	\label{eq:phi}
\end{equation}
where $\phi_0$, $\nu_0$ and $\dot\nu_0$ are the phase, frequency and first frequency derivative of the pulsar at time $t_0$, respectively. Fitting the measured phase drift $\psi_{\rm obs}(t)$ with equations~(\ref{eq:psi}) and~(\ref{eq:phi}), we obtained the timing solution of the Iqueye December observations shown in Fig. \ref{fig:Iqueye_drift} and Table \ref{tab:drift_fitpars}, which is in perfect agreement with that reported in the Second \fermi-LAT Catalog of Gamma-ray Pulsars (2PC; \citealt{Abdo2013}).

\begin{table}
\caption{Timing solution of the 2009 December run. Errors represent the 1$\sigma$ statistical uncertainties.}
\label{tab:drift_fitpars}
\centering
\begin{tabular}{l l}
\hline\hline
\multicolumn{2}{c}{Timing solution of the Iqueye 2009 December}	\\
\hline
$t_0$		& MJD 55182	\\
$\phi_0$		& $-0.4495 \pm 0.0040$	\\
$\nu_0$		& $11.18981699 \pm 5\times10^{-8}$ Hz	\\
$\dot\nu_0$	& $(-1.35 \pm 0.13) \times 10^{-11}$ Hz$^2$	\\
$P_{\rm init}$	& $0.08936694$ s	\\
\hline
\end{tabular}
\end{table}

\subsection{Iqueye Pulse profile}\label{sec:3.2}
The limited number of Iqueye observations (one night in January and three consecutive nights in December) does not allow us to phase connect the phase of the Vela pulsar over the entire year 2009. Therefore, we folded the whole Iqueye dataset (Obs. 4, 7, 12, 14 and 15) using the accurate timing solution obtained by the \fermi-LAT team within the framework of the 2PC \fermi~catalog. The 2PC timing solution\footnote{Available here: \url{https://fermi.gsfc.nasa.gov/ssc/data/access/lat/ephems/}} is based on the rotational ephemerides computed using the 1.4 GHz radio data collected at the \textit{Parkes} radio telescope. It covers more than 4 years from MJD 54207.2 to MJD 55793.85 and contains ten frequency derivatives and fifteen sinusoid WAVE terms (see \citealt{Hobbs2006} for details). It also includes all the information needed to perform an absolute phase alignment with the radio pulse.

We compared the 2PC \fermi-LAT ephemerides with another publicly available ephemerides, which covers the dates of the Iqueye observing runs. These ephemerides have been computed using the times of arrival of the electromagnetic pulses measured with the Hobart radio telescope in Tasmania (Australia) and the Hartebeesthoek radio telescope in South Africa during the VSR2 run of the Virgo gravitational wave detector, which began on 2009 July 7th and ended on 2010 January 8th \citep{Abadie2011}. The timing solution contains two frequency derivatives. We found that it is accurate enough for performing the folding of the Iqueye data. However, the timing model from \citet{Abadie2011} has a significant difference in absolute phase (up to $\sim$0.4) from the 2PC timing model.

We then folded the Iqueye data with the 2PC \fermi-LAT ephemerides using different numbers of phase bins and found that 128 bins provide the best compromise between the error on the count rate per bin and the resolution in phase. The total folded profile was obtained summing in phase those of the single observations (Obs. 4, 7, 12, 14 and 15). The resulting pulse shape is shown in Fig. \ref{fig:Iqueye_Fermi_VelaLC} and is characterized by the presence of at least four distinct peaks.

In order to obtain an accurate analytical representation of the Vela pulse profile, we fitted it with the sum of six Gaussian functions plus a constant:
\begin{equation}
	f_{\rm opt}(\phi) = \sum_{i=1}^6 k_i\,\frac{1}{\sqrt{2\pi}\sigma_i}\,\exp\left[-\frac{(\phi-\mu_i)^2}{2\sigma_i^2}\right] + C \, .
	\label{eq:opt_template}
\end{equation}
The values of the best-fitting parameters are listed in Table \ref{tab:Iqueye_Vela_templates}, while the resulting analytical template is shown in Fig. \ref{fig:Iqueye_Fermi_VelaLC}. The reduced $\chi^2$ of the fit is 0.977. Three of the peaks in the pulse profile (P1s, P3 and P4) are fitted with single Gaussians centered at phases 0.256, 0.841, 1.0, respectively. Peak P2s at phase $\sim$0.47 is fitted with the sum of two Gaussian components (P2s-1 and P2s-2). Another broad Gaussian at phase 0.56 is required to reproduce the complex background.

\begin{table}
\caption{Best-fitting parameters of function $f_{\rm opt}$ (equation \ref{eq:opt_template}) adopted to fit the Iqueye pulse profile of the Vela pulsar. Errors represent the 1$\sigma$ statistical uncertainties.}
\label{tab:Iqueye_Vela_templates}
\centering
\begin{tabular}{c c c c}
\hline\hline
$i$	& $\mu_i$	& $\sigma_i$	& $k_i$	\\
\hline
1	& $0.256 \pm 0.004$	& $0.049 \pm 0.004$	& $520 \pm 50$	\\
2	& $0.451 \pm 0.003$	& $0.012 \pm 0.004$	& $100 \pm 60$	\\
3	& $0.497 \pm 0.007$	& $0.031 \pm 0.007$	& $350 \pm 90$	\\
4	& $0.56 \pm 0.04$	& $0.14 \pm 0.04$	& $400 \pm 100$	\\
5	& $0.841\pm 0.004$	& $0.016 \pm 0.005$	& $80 \pm 20$	\\
6	& $1.000 \pm 0.002$	& $0.012 \pm 0.002$	& $90 \pm 20$	\\
\noalign{\vskip 1mm} 
\multicolumn{4}{l}{$C = 487100\pm 200$}	\\
\hline
\end{tabular}
\end{table}

To estimate the significance of the various peaks we compared the best-fitting model $f_{\rm opt}$ with one in which the amplitude of each component was reduced by a factor of 5. We computed the F-statistics and the corresponding probability by comparing the $\chi^2$ of the fits with these two models, obtaining the significance of each component in Gaussian sigmas. For the two prominent peaks P1s and P2s the significance is clearly very high, while that of peaks P3 and P4 is 3.9$\sigma$ and 5.5$\sigma$, respectively. We also estimated that the significance of the narrower peak P2s-1 at phase 0.451 is 5.7$\sigma$.

In order to double check the robustness of our results, we repeated all the analysis folding the pulse profiles using 64 bins in phase and obtained similar results (the significance of peaks P2s-1, P3 and P4 is 5.1$\sigma$, 3.6$\sigma$ and 4.8$\sigma$, respectively).

\subsection{\fermi-LAT Pulse profile}\label{sec:3.3}
To perform an accurate comparison of the phase of the optical peaks to that at other wavelengths, we used as reference the gamma-rays. We then analyzed the \fermi-LAT data with the \fermi~Science Tools software package v11r5p3\footnote{\url{https://fermi.gsfc.nasa.gov/ssc/data/analysis/software/}.}, and obtained the high-energy pulse profile of the Vela pulsar using a technique similar to one described in Sect. \ref{sec:3.2}.

We used more than 3 yrs of available data accumulated by the \fermi~satellite during the time interval from MJD 54682.66 to MJD 55793.85, fully covered by the 2PC  timing solution. We extracted events with energies between 30 MeV and 300 GeV. To improve the signal to noise ratio we used an energy-dependent extraction region selecting only events within an angle $\theta<\max[1.6-3\log_{10}(E_{\rm GeV}), 1.3]$ deg from the Vela pulsar position\footnote{The equatorial coordinates of the Vela pulsar are taken from the 2PC ephemerides: $\alpha=128.835839242^\circ$, $\delta=-45.176336389^\circ$.}, where $E_{\rm GeV}$ is the energy of the event measured in GeV (see \citealt{Abdo2009_4, Abdo2010_3}). We excluded events detected during the periods when the Vela pulsar is viewed at zenith angles $>$90$^\circ$ in order to minimize the contamination from the Earth limb. After selecting the events and determining the good time intervals of the \fermi-LAT data we performed a barycentric correction of the photon arrival times using the tool \texttt{gtbary}.

\begin{figure*}
	\centering
	\includegraphics[width=0.85\textwidth]{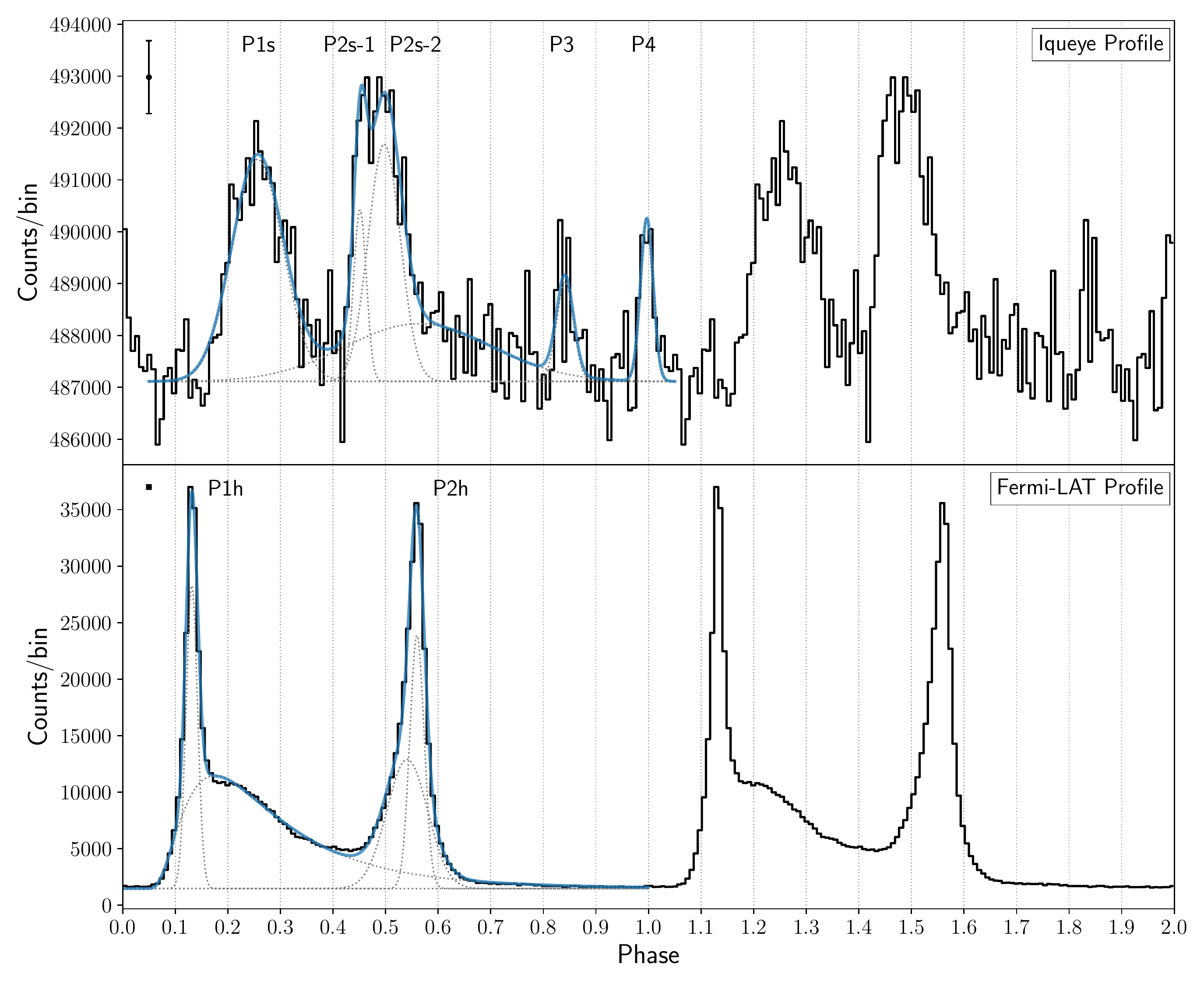} \\
	\caption{Two periods of the Vela pulsar pulse profile from the Iqueye optical observations taken in 2009 (\textit{top panel}) and the \fermi-LAT gamma-ray observations taken during 3 yrs of its operation (MJD 54682.66--55793.85, \textit{bottom panel}). The blue line in each plot shows the best-fitting analytical function which reproduces the profile. Individual components of these functions are also shown with gray dashed lines. Phase 0 corresponds to the radio peak. The average 1$\sigma$ error bar is shown on the top left.}
	\label{fig:Iqueye_Fermi_VelaLC}
\end{figure*}

We calculated the phase of each photon and folded the \fermi-LAT data using the 2PC timing solution. The final gamma-ray pulse profile of the Vela pulsar is shown in Fig. \ref{fig:Iqueye_Fermi_VelaLC}. As for the Iqueye profile, we used 128 phase bins. We then fitted the gamma-ray pulse profile with an analytical function $f_{\gamma}$ which is the sum of tree Gaussians, a log-normal function and a constant\footnote{We used Gaussian components to fit the gamma-ray peaks and not asymmetric Lorentzian functions (as e.g. in \citealt{Abdo2010_3}) to avoid additional systematic uncertainties when comparing the Iqueye and \fermi-LAT peaks profiles. We also tried to replace the log-normal component with a fourth Gaussian function, but it worsened the fit.}:

\begin{multline}
	f_{\gamma}(\phi) =  \sum_{i=1}^3 k_i\,\frac{1}{\sqrt{2\pi}\sigma_i}\,\exp\left[-\frac{(\phi-\mu_i)^2}{2\sigma_i^2}\right] + \\
+ k_4\,\frac{1}{(\phi-\phi_0)\,\sqrt{2\pi}\sigma_4}\,\exp\left[-\frac{(\ln(\phi-\phi_0) - \mu_4)^2}{2\sigma_4^2}\right] +  C.
	\label{eq:gamma_template}
\end{multline}
The best-fitting values of the parameters are listed in Table \ref{tab:Fermi_Vela_templates}. The reduced $\chi^2$ value of the fit is rather high (16.4) because of the small error bars of the \fermi-LAT profile. However, the two main gamma-ray peaks P1h and P2h are well reproduced for our purposes.

\begin{table}
\caption{Best-fitting parameters of the function $f_{\gamma}$ (equation \ref{eq:gamma_template}) adopted to fit the \fermi-LAT pulse profile of the Vela pulsar. Errors represent the 1$\sigma$ statistical uncertainties.}
\label{tab:Fermi_Vela_templates}
\centering
\begin{tabular}{c c c c c}
\hline\hline
$i$	& $\mu_i$	& $\sigma_i$	& $k_i$ & $\phi_0$	\\
\hline
1	& $0.1315 \pm 0.0002$	& $0.0109 \pm 0.0002$	& $730 \pm 20$	& $-$	\\
2	& $0.5411 \pm 0.0009$	& $0.0387 \pm 0.0008$	& $1110 \pm 30$	& $-$	\\
3	& $0.5597 \pm 0.0003$	& $0.0137\pm 0.0004$	& $770 \pm 30$	& $-$	\\
4	& $\ln(0.208 \pm 0.004)$	& $0.66 \pm 0.02$	& $2770 \pm 60$	& $0.040 \pm 0.004$	\\
\noalign{\vskip 1mm} 
\multicolumn{5}{l}{$C = 1460 \pm 40$}	\\
\hline
\end{tabular}
\end{table}

\begin{table}
\caption{Peak positions of the Iqueye and \fermi-LAT pulse profiles. Errors represent the estimated 1$\sigma$ statistical uncertainties (see text for details).}
\label{tab:Iqueye_Fermi_Vela_peaks}
\centering
\begin{tabular}{l l}
\hline\hline
\multicolumn{2}{c}{Iqueye pulse profile}	\\
\hline
P1s 		&  $0.257 \pm 0.004$	\\
P2s-1 	&  $0.455^{+0.005}_{-0.003}$	\\
\noalign{\vskip 1mm} 
P2s-2 	&  $0.498^{+0.004}_{-0.007}$	\\
P3 		&  $0.841 \pm 0.005$	\\
P4 		&  $0.997 \pm 0.002$	\\
\noalign{\vskip 1mm} 
\multicolumn{2}{c}{\fermi-LAT pulse profile}	\\
\noalign{\vskip 0.6mm} 
P1h 		&  $0.1319 \pm 0.0002$	\\
P2h 		&  $0.5586 \pm 0.0003$	\\
\hline
\end{tabular}
\end{table}

\subsection{Relative timing of the radio-optical-gamma-ray peaks}\label{sec:3.4}
As the Iqueye pulse profile contains multiple overlapping components, we determined the centroid of each peak calculating the local maxima of the function $f_{\rm opt}$. To this end, we computed the zeros of the derivative $f'_{\rm opt} \equiv df_{\rm opt}/d\phi$. Uncertainties of the peak positions were estimated from the difference of the zeros of the functions $f'_{\rm opt}+\delta f'_{\rm opt}$ and $f'_{\rm opt}-\delta f'_{\rm opt}$, where $\delta f'_{\rm opt}$ is obtained propagating the errors of the fitted parameters $p_i$ on $f'_{\rm opt}$:
\begin{equation}
	\delta f'_{\rm opt} = \sqrt{\sum_i \sum_j \left(\frac{\partial f'_{\rm opt}}{\partial p_i}\right) \left(\frac{\partial f'_{\rm opt}}{\partial p_j}\right) \sigma_{ij}}.
	\label{eq:opt_df}
\end{equation}
$\sigma_{ij}$ are elements of the covariance matrix of the fit. The final positions of the peaks of the Iqueye optical pulse profile determined in this way are listed in Table \ref{tab:Iqueye_Fermi_Vela_peaks}. The positions of the peaks P1h and P2h of the \fermi-LAT gamma-ray pulse profile and their corresponding uncertainties have been calculated in the same way and are also listed in Table \ref{tab:Iqueye_Fermi_Vela_peaks}.

The quality of the Iqueye pulse shape allows us to determine the relative time shift between the narrower optical and gamma-ray/radio peaks with an accuracy of a fraction of a millisecond. The gamma-ray peak P1h is leading the optical peak P1s by $0.125 \pm 0.004$ ($11.2 \pm 0.4$ ms). The high-energy peak P2h is trailing both optical peaks P2s-1 and P2s-2 by $0.104^{+0.005}_{-0.003}$ and $0.061^{+0.003}_{-0.007}$, respectively. Peak P3 is leading the radio peak by $0.159 \pm 0.005$ ($14.2 \pm 0.4$ ms), whereas peak P4 turns out to be aligned in phase with the radio peak within the 3$\sigma$ errors.

The dominant source of the uncertainties are fitting errors on the parameters of the functions $f_{\rm opt}$ and $f_{\rm \gamma}$. The broadening of the peaks caused by uncertainties of the 2PC \fermi-LAT timing solution has a small influence. To confirm this we folded both the Iqueye and \fermi-LAT data with two timing solutions that, considering the uncertainties on the 2PC ephemerides, extremize the deviations from the mean timing solution. The differences between the peak positions determined by means of these two timing solutions are smaller than corresponding fitting errors and, therefore, can be neglected.

\section{DISCUSSION AND CONCLUSIONS}\label{sec:4}
The aim of this work is to present the analysis of the optical data of the Vela pulsar obtained with the Iqueye instrument at the ESO NTT telescope. The observations collected with Iqueye were used to produce for the first time an independent optical timing solution. We also obtained the most detailed optical pulse profile of the Vela pulsar available to date.

The pulse shape of the Vela pulsar detected with Iqueye has a better resolution and shows a finer structure than those obtained in previous optical investigations (e.g. \citealt{Manchester1980,Gouiffes1998}). It shows new features such as the double-peak structure of peak P2s and a significant peak P3, that were not clearly identified in the optical band before. The overall quality of the profile is comparable to that obtained in the UV band by \citet{Romani2005} using the \hst~STIS (Space Telescope Imaging Spectrograph). 

We determined the phase of each optical peak with an accuracy of a fraction of a millisecond relative to the peaks in the radio and gamma-ray bands. Such accurate measurements can be used in the future for testing theoretical models which describe the location of the pulsar emitting regions. Here we simly note that, although within the formal errors the two peaks are aligned, in fact there is some hint that the optical peak P4 may be leading the radio one by a few hundred microseconds (0.003 in phase). This would be perfectly consistent with the shift measured in the Crab pulsar \citep{Zampieri2014}.

\begin{figure*}
	\centering
	\includegraphics[width=0.75\textwidth]{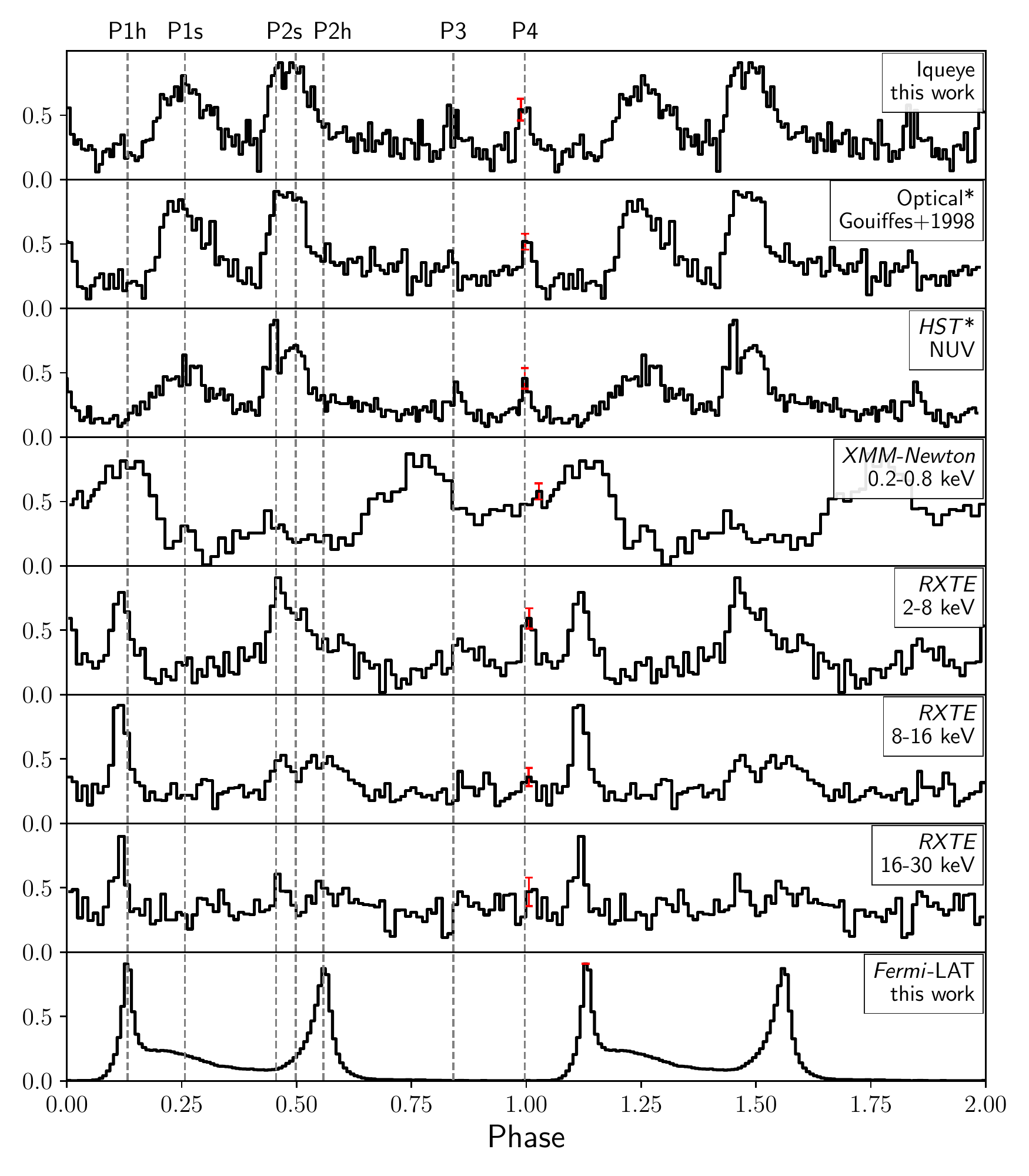} \\
	\caption{Normalized pulse profiles of the Vela pulsar in different energy bands. Optical, near UV and X-rays profiles are adopted from \citet{Gouiffes1998}, \citet{Romani2005}, \citet{Kuiper2015} and \citet{Harding2002}, respectively. Pulse profiles from \citet{Gouiffes1998} and \citet{Romani2005} (marked with asterisks) are shifted so as to align them with the Iqueye profile (see text for details). Phase 0 corresponds to the radio peak. The vertical dotted lines mark the positions of different peaks. The average 1$\sigma$ error bar for each profile is shown in red.}
	\label{fig:Comp_LCs}
\end{figure*}

In Fig. \ref{fig:Comp_LCs} we compare the shape of the Iqueye and \fermi-LAT pulse profiles obtained here with those at other wavelengths. The optical, UV and X-rays profiles are taken from \citet{Gouiffes1998}, \citet{Romani2005} and \citet{Harding2002, Kuiper2015}, respectively. As the accuracy on the absolute phase measurement of \citet{Gouiffes1998} is not known and the absolute timing precision of the \hst~STIS data in \citet{Romani2005} is limited to 10 ms, in Fig. \ref{fig:Comp_LCs} we shifted the corresponding pulse profiles as to align them with the more accurate profile obtained with Iqueye\footnote{This shift was calculated from the average of the difference between the position of the Iqueye optical peaks P1s, P2s-1, P2s-2, P3 and P4 (Table~\ref{tab:Iqueye_Fermi_Vela_peaks}) and that of the \citet{Gouiffes1998} and \citet{Romani2005} pulse profiles, the latter obtained by fitting each peak with a Gaussian function.}.

The multi-wavelength pulse profile of the Vela pulsar shows at least five components. Starting from phase zero, at the radio peak, and moving to larger phases, we see first some emission at low energies which is difficult to disentangle from that of the underlying background. Then, at phase $\sim$0.1, the line of sight suddenly enters in the narrow bright gamma-ray peak P1h emission region, where X-ray emission is also significant \citep{Harding2002, Kuiper2015} and optical emission is only marginal. At larger phases, the line of sight passes through the optical peaks P1s and P2s. Peak P2s is well structured in the optical and UV bands. There is a narrow peak on its edge (denoted as P2s-1), which is aligned with one in the X-rays. In gamma-rays we see a region of continuous moderate bridge emission, which changes morphology with increasing energy \citep{Abdo2009_4, Abdo2010_3}. This region terminates sharply at phase 0.6, where the second gamma-ray/X-ray peak P2h is located. From phase 0.6 through phase 1, gamma-ray emission is not present and no other prominent emission features are visible in the other spectral bands, with the exception of peak P3 and a broad soft X-ray peak at phase 0.8 (whose tail extends up to peak P1h, forming a sort of bridge emission, \citealt{Kuiper2015}). At phase 1.0 the line of sight crosses the radio peak, which is aligned with peak P4 visible in the optical, UV and X-ray bands.

We suggest that the optical emission originates from three different regions: optical peaks P1s and P2s-1 are likely to originate near the X-ray and gamma-ray emission regions, whereas optical peaks P3 and P4 may form closer to the soft X-ray and radio emission regions, respectively. The optical/hard X-ray/gamma-ray emission region may be located along magnetic field lines originating from a single magnetic pole (\citealt{Manchester1977,Manchester1980}) or neighboring magnetic field lines, while the radio/optical and soft X-ray/optical emission originate  far from them. Indeed, none of main X-ray/gamma-ray peaks is in phase with the radio peak at phase 0. Such a complex emission pattern could be produced by electrons accelerated to higher and higher energies as they travel along a bundle of field lines that opens more and more farther out in the pulsar magnetosphere. This scenario is consistent with models that favor a high-altitude origin of the gamma-ray emission (see e.g. \citealt{Romani1995,Bai2010_1,Bai2010_2,Rudak2017arx}).

Future optical observations could allow us to achieve even better statistics in determining the pulse profile of the Vela pulsar and its fine structure, which turns out to be possible only for a handful of pulsars. This is needed to complement the information from other high energy bands and improve our understanding of the complex emission and magnetic field geometry of the Vela pulsar.

\section*{Acknowledgements}
We would like to thank an anonymous referee for his/her constructive and insightful comments that helped us improving the paper. We also thank Tommaso Occhipinti, Ivan Capraro, Andrea di Paola and Paolo Zoccarato for their crucial contribution to the development of the acquisition system of Iqueye and for their support with the observations and data collection at the telescope. AP acknowledges support from a Netherlands Organization for Scientific Research (NWO) Vidi fellowship. This work is based on observations made with the ESO NTT telescope at the La Silla Paranal Observatory under programmes IDs 082.D-0382 and 084.D-0328(A). This research has been partly supported by the University of Padova under the Quantum Future Strategic Project, by the Italian Ministry of University MIUR through the programme PRIN 2006, by the Project of Excellence 2006 Fondazione CARIPARO, and by INAF-Astronomical Observatory of Padova. This research made use also of the following \texttt{PYTHON} packages: \texttt{MATPLOTLIB} (Hunter 2007), \texttt{NumPy} (van der Walt et al. 2011) and \texttt{Astropy} (Astropy Collaboration et al. 2013).

\bibliographystyle{mnras} 

\appendix
\section{Vela Pulsar Period Search in the Iqueye Data}\label{app:1}
We carried out an independent search for pulsations on the optical data to test the sensitivity and accuracy of our instrument to the pulsar signal. We then calculated power density spectra (PDS) of all light curves using different time bins (from 1 ms to 10 ms) with the XRONOS task \texttt{powspec} \citep{Stella1992}. The time bin was varied in order to identify possible spurious low-significance peaks induced by the noise. The PDS are Leahy normalized \citep{Leahy1983}. For an optimal detection of periodic signals we computed a single PDS for the entire length of each observation (frequency resolution $\approx$10$^{-4}$ Hz). 

One observation (Obs. 14) shows a clear peak at a significance $\sim$9$\sigma$ single trial ($7 \sigma$ for 539915 independent trials) in the power spectrum. The frequency of this peak is 11.1898 Hz. Other two observations show peaks at 11.1898 Hz with a significance of $\sim$6$\sigma$ single trial (Obs. 12 and 15).

\begin{figure}
\centering
	\includegraphics[width=0.5\textwidth]{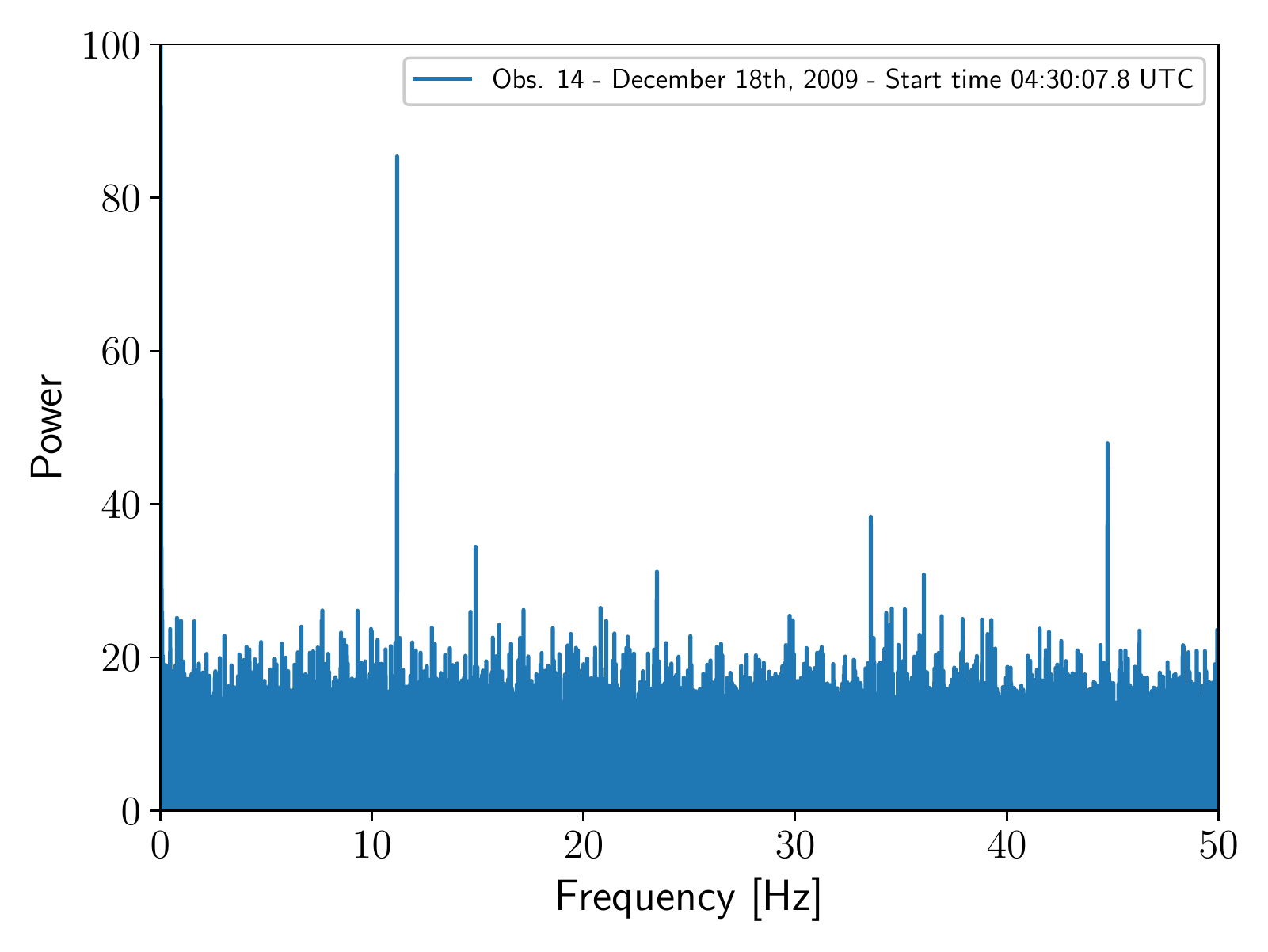}
	\caption{Leahy normalized power spectrum of Obs. 14 (see Table 1). The time bin is 0.01 s. The resolution in frequency is 10$^{-4}$ Hz. }
	\label{fig:pdsuk}
\end{figure}

The peak at the lowest frequency corresponds to the fundamental frequency associated to the rotation of the pulsar and provides its period of rotation, $p=1/{\nu} = 0.089367$ s. Interestingly, the frequencies at 33.5694 Hz and 44.7592 Hz detected in the power spectrum of Obs. 14 correspond to the third and fourth harmonics of the fundamental and are present (though not always) also when varying the time bin. Furthermore, as it can be seen in the optical pulse profile, the distance in phase between the two main peaks is approximately 1/4 in phase, which is consistent with the presence of a significant fourth harmonic component in the power spectrum.

\begin{figure}
\centering
	\includegraphics[width=0.5\textwidth]{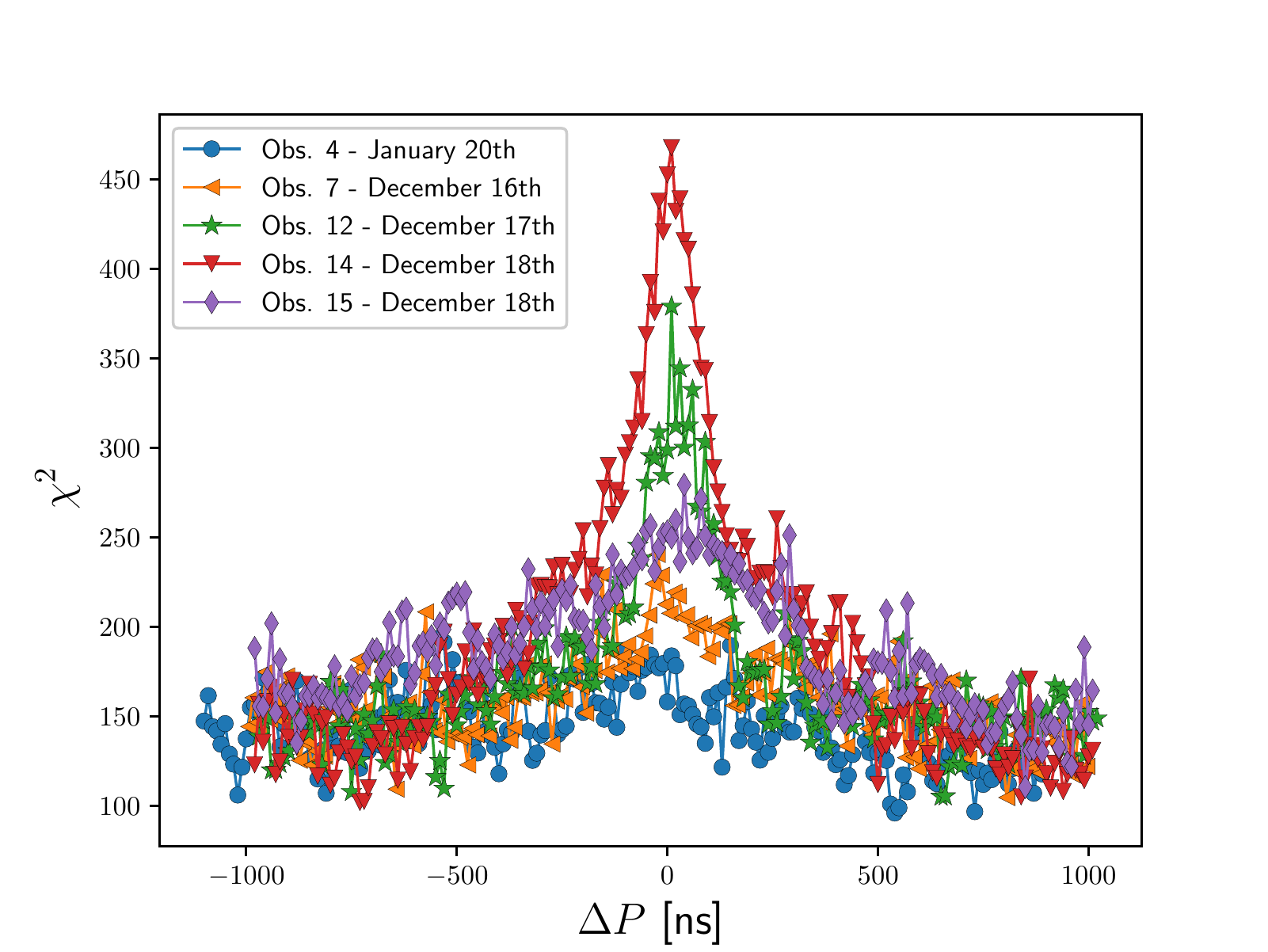}
	\caption{Distribution of the ${\chi}^2$ for the Vela pulsar light curves, folded with different rotational periods. Each curve is shifted with the corresponding value of the rotational period reported in Table \ref{tab:periodiIqueye}. Data is barycentered in \texttt{TEMPO2} mode.} 
	\label{fig:bestperiod}
\end{figure}

We then performed a refined search with the folding technique using the task \texttt{efsearch} in XRONOS. The final rotational periods obtained for these observations (Obs. 12, 14 and 15) and the maximum values of the ${\chi}^2$ are reported in Table \ref{tab:periodiIqueye} (for data barycentered with \texttt{TEMPO2}). The corresponding ${\chi}^2$ curves are shown in Fig. \ref{fig:bestperiod}.
The error on the rotational period was estimated by fitting the peak of the distribution of the ${\chi^2}$ with a Gaussian\footnote{The time resolution for the period search is $<$100 ns. For larger resolutions the peak is not sufficiently well defined.}. The dispersion of the Gaussian provides a conservative estimate of the error, which turns out to be $\sim$20 ns.

We used the 2PC \fermi-LAT ephemerides to search for pulsations in the optical observations with no significant signal in the power spectrum, including those taken in January. The folding technique allowed us to find a weak signal in other two observations, Obs. 4 in January and Obs. 7 in December. The corresponding rotational periods and ${\chi}^2$ curves are also shown in Table \ref{tab:periodiIqueye} and Fig. \ref{fig:bestperiod}. 

\begin{table}
\caption{Periods of the Vela pulsar obtained with Iqueye. Data is barycentered in \texttt{TEMPO2} mode. The last digit in brackets is the 1$\sigma$ statistical error.}
\label{tab:periodiIqueye}
\begin{tabular}{ccccc}
\hline
\hline
\noalign{\smallskip}
Obs.	& Start time	& Frequency	& Period	& ${{\chi}^2_{\rm max}}^a$	\\
	& [MJD] 		& [Hz]		& [s] 	&	\\
\hline
\noalign{\smallskip}
4   & 54851.230534 & 11.19033(5) &  0.08936289(4) & 190	\\
7   & 55181.168850 & 11.18982(3) &  0.08936695(2) & 240	\\
12 & 55182.193145 & 11.18982(1) & 0.08936694(8) & 379	\\
14 & 55183.187590 & 11.18982(1) & 0.08936698(9) & 468	\\ 
15 & 55183.315835 & 11.18980(3) & 0.08936710(2) & 279	\\ 
\noalign{\smallskip}
\hline
\noalign{\smallskip}
\end{tabular}
{\smallskip}
\begin{minipage}{15 cm}
{$^a$ Maximum value of the $\chi^2$ reported in Fig. \ref{fig:bestperiod}. \\
}
\end{minipage}
\end{table}

\bsp	
\label{lastpage}
\end{document}